\documentclass[12pt]{iopart}
\usepackage{graphicx}  % needed for figures
\usepackage{bm}        % for math
\usepackage{amssymb}   % for math
\usepackage{ulem}
\usepackage{enumerate}
\begin{document}
	\title{Long-range interaction induced collective dynamical behaviors}
	\author{K. Sathiyadevi$^{1}$,  V. K. Chandrasekar$^{1}$,  D. V. Senthilkumar$^2$ and M. Lakshmanan$^{3}$}
	\address{$^1$Centre for Nonlinear Science \& Engineering, School of Electrical \& Electronics Engineering, SASTRA Deemed University, Thanjavur -613 401, Tamil Nadu, India. \\$^2$School of Physics, Indian Institute of Science Education and Research, Thiruvananthapuram-695551, India. \\ 
		$^3$Centre for Nonlinear Dynamics, School of Physics, Bharathidasan University, Tiruchirappalli - 620 024, Tamil Nadu, India. }
	\ead{chandru25nld@gmail.com}
	\vspace{10pt}
	%\begin{indented}
	%\item[]August 2017
	%\end{indented}
	%
	\begin{abstract}
		Long-range interacting systems are omnipresent in nature. We investigate here the collective dynamical behavior in a long-range interacting system consisting of coupled Stuart-Landau limit cycle oscillators.  In particular, we analyze the impact of a repulsive coupling along with a symmetry breaking coupling. We report that the addition of repulsive coupling  of sufficient strength can induce a swing of the synchronized state which will start disappearing with increasing disorder as a function of the repulsive coupling.     We also deduce analytical stability conditions for the oscillatory states including synchronized state, solitary state, two-cluster  state as well as  oscillation death state. Finally,  we have also analyzed the effect of power-law exponent  on the observed dynamical states. 
	\end{abstract}
	%\maketitle
	
	\section{Introduction}{\label{int}}

	\par  Complex dynamical systems exhibit a variety of dynamical patterns through distinct interesting transitions due to the nature of coupling  architectures.  Among the various coupling geometries,  investigations on coupled networks with global (all to all) coupling have become an active area of research in recent times, because of their prevalence  in many natural systems \cite{glo1, glo2,  appl_1, rev2_glob}. Moreover, such a type of network architecture is often used to study the collective properties of chemical reactions \cite{chemic1, chimec3}, physical systems \cite{phys1, phys2, phys3, phys4}, electronic devices \cite{elect} and biological systems \cite{bio, bio1}.   Globally coupled dynamical systems exhibit diverse dynamical patterns such as synchronization, clustering and  many more states including one of the emergent hybrid dynamical behavior called  chimera state \cite{scho_glob,clust1, clust2}.   In addition, the network of coupled dynamical systems exhibit different transitions with coexisting distinct dynamical states.  But still the  underlying  mechanism behind such  transitions, and multistabilities among the distinct dynamical behaviors, remain unclear.   Further it is also known that globally (mean-field) coupled systems can act as  long-range interacting systems  under certain limitations \cite{lr_glob}.   Studies on long-range interaction have also been an active area of research over the last several decades \cite{lr_lec,s.gupta1,s.gupta2,s.gupta3}. Generally, any complex interaction decaying as a power law at large distances,  without any specific length,  can be considered as long range interaction.	One can find  such systems in astrophysics, plasma physics, atomic physics, nuclear physics and hydrodynamics \cite{lr1, astro1, plasma, hyd}. Due to the widespread nature of long-range interacting systems, it is essential to understand  the dynamical behavior of an   ensemble of oscillators with this kind of interactions.  The effect of such interactions can be well understood  with  mean-field (global) coupling \cite{ata1, ata2}.  
	
	\par  On the other hand,  studies on synchronization and its control have been receiving much attention due to their potential  applications in  many physical, biological, social and  even in man-made systems \cite{sync}. Particularly, a large scale neuronal synchronization causes several brain disorders such as Parkinson's disease, tremor and epilepsy \cite{syn0, pop_syn1, syn2, awa_pras}.  Though many investigations have been made to understand their intricacies in different systems, still it demands a deep knowledge to control the mechanism of synchronized states.  Therefore a continuous effort is still on to explore many aspects associated with the phenomenon of synchronization \cite{syn_dana, syn_kapi}.   Out of numerous investigations on synchronized states,  Daido and Nakanishi  had found  a novel interesting behavior called a  ``swing of the synchronized state" in a large population of globally coupled Stuart-Landau oscillators where the inhomogeneity among the synchronized regions emerges by introducing  diffusion  in a globally coupled system in the  presence of nonisochronicity parameter.  The investigation by Daido and Nakanishi revealed that the synchronized   state is destabilized while increasing the   coupling strength which  again gets stabilized upon increasing  the coupling strength further. The destabilization of stable  synchronized state and again re-stabilization of  destabilized synchronized state  is called  the swing of the synchronized state \cite{daido_anal1, daido_anal2}. Later, such  an interesting dynamical  behavior was also found  in a  globally coupled system of Stuart-Landau oscillators with symmetry breaking coupling in  the  presence of nonisochronicity parameter by Premalatha~\textit{et~al} \cite{main_prema}. These authors reported that the symmetry breaking coupling facilitates the widening of   the dynamical regime by twice as compared with the symmetry preserving coupling, and an increase of disorderliness when increasing the nonisochronicity parameter. Besides the above,  recent investigations revealed that the trade-off between repulsive and  attractive couplings  can exhibit diverse collective dynamical states, including  spontaneous symmetry breaking, along with distinct multistablities  in a single  network \cite{sathya1,sathya2,skdana}.
	In the present study, we explore \textit{whether the addition of repulsive coupling  along with symmetry breaking coupling can induce the swing by mechanism and, if so, what is its influence on the degree of disorder among the coupled systems}. 
	Indeed,  we find  the emergence of  swing of the synchronized state while increasing the strength of the repulsive coupling up to a  sufficient value  beyond which it  will disappear with increasing disorderliness at   strong   repulsive coupling. 
	\par In this report, we investigate the  dynamical transitions in a network of long-range  coupled Stuart-Landau oscillators with the addition of repulsive link in the symmetry breaking coupling. Initially, we show the emergence of a swing of  synchronized state for  sufficient strength  of the  repulsive coupling in the absence of power-law exponents and  increasing   disorder with the  increase of the  strength of the repulsive coupling.  Further, the swing of the synchronized state disappears at strong repulsive coupling strength and the systems attain a completely disordered state. In the swing of the synchronized state the dynamical transitions are observed through the synchronized state to solitary state and cluster state and finally  re-emergence of the synchronized state.  We have also deduced the analytical stability conditions for the oscillatory states of synchronized states, solitary state and cluster states.  Finally, we obtain the stability conditions for oscillation death state. We also identify the dynamical transitions for  different  power-law exponents and report that the system becomes completely desynchronized  when increasing the power-law exponent to appropriate higher values.
	\par The rest of the paper is organized as follows.  In Sec.~\ref{mod}, we introduce our model of long-range coupled Stuart-Landau oscillators.   The numerical results for swing by mechanism and global behavior of the coupled system are discussed in the absence of power-law exponents (i.e. the system is all-to-all coupled)  in  Sec.~\ref{numer}.  The corresponding analytical results are discussed in Sec.~\ref{anal}. We analyze the dynamical transitions by varying the power-law  exponents in Sec.~\ref{plc}.  Finally, we summarize our findings in Sec.~\ref{con}.
	\section{The Model}\label{mod}
	\par  Stuart-Landau limit cycle oscillator is a general, paradigmatic model, which can be used to model a variety of nonlinear dynamical systems near the Hopf bifurcation point \cite{stuart1, stuart2, stuart3, stu_ap}.  We consider a one-dimensional ring  network of  coupled  Stuart-Landau oscillators and  investigate its dynamical behavior by including a  repulsive link along with a  symmetry breaking coupling. The proposed coupling is a symmetry breaking coupling which breaks the rotational symmetry of the coupled system explicitly.   The governing equations of the network can be represented as
	\begin{eqnarray}
	\fl \quad \dot{ z}_{j}=f(z_j)+ \frac{\kappa}{\eta (\alpha)} \, \bigg[ \sum_{j\ne k}^{}\frac{1}{|j-k|^\alpha} \big(Re({z_k}-{  z}_{j}) -i\, q \,Im ({z_k}-{  z}_{j})\big)\bigg], \,\,j, k = 1, 2, ..., N,
	\label{model1}
	\end{eqnarray}
	where $f(z_j)=(\lambda+i\omega)z_j-(1-ic)|z_j|^2 z_j$, the state variables being $z_j=x_j+i y_j \in C$. % $i = 1,2,...,N$,   
	In the absence of coupling the considered  system preserves symmetry (rotational invariance), that is the system equation remains invariant under the transformation $z_j\rightarrow z_je^{i\theta}$. In contrast, the rotational invariance is not preserved upon introducing the couplings  and hence these types of couplings are known as symmetry breaking couplings \cite{sbre1, sbre2}. Here $N$ is the total number of the oscillators in the network and  $\eta(\alpha)=\sum_{k\ne j}^{N}{|j-k|^{-\alpha}}$ is  the normalization constant. When  the strength of the interaction lies between  $0 \le \alpha \le 1$,  the interaction  is  said to be of long range. On the otherhand $\alpha > 1$ is considered as the short range.  The edge of long-range interaction is achieved through $\alpha=0$, when  the system becomes mean-field (globally) coupled whereas $\alpha\rightarrow\infty$ indicates that systems are  coupled only to the  nearest neighbors \cite{lr_range1, lr_range2, lr_range3, lr_range4}.    Here,  $\kappa= ((N-1)/N)\epsilon$, where $\epsilon$ is the coupling strength and $q$ is the  strength of the repulsive coupling.  $\lambda$ and $\omega$ correspond to  the bifurcation parameter and the natural frequency of the system, respectively.    c is the nonisochronicity parameter. Throughout the manuscript, the parameter values are fixed as $\lambda=1.0$, $\omega=2.0$, $c=2.5$  and $N=100$. The initial state for the state  variables  $x_i$ and $y_i$  are chosen  randomly between -1 to 1 and   Runge-Kutta fourth order integration scheme is used  for all our simulations with a time step 0.01. 
	\section{ Dynamical transitions for power-law exponent $\alpha=0$}\label{numer}
	\subsection{ Swing of synchronized states }
	\begin{figure*}[h]
		\centering
		\hspace{-0.1cm}
		\includegraphics[width=16.0cm]{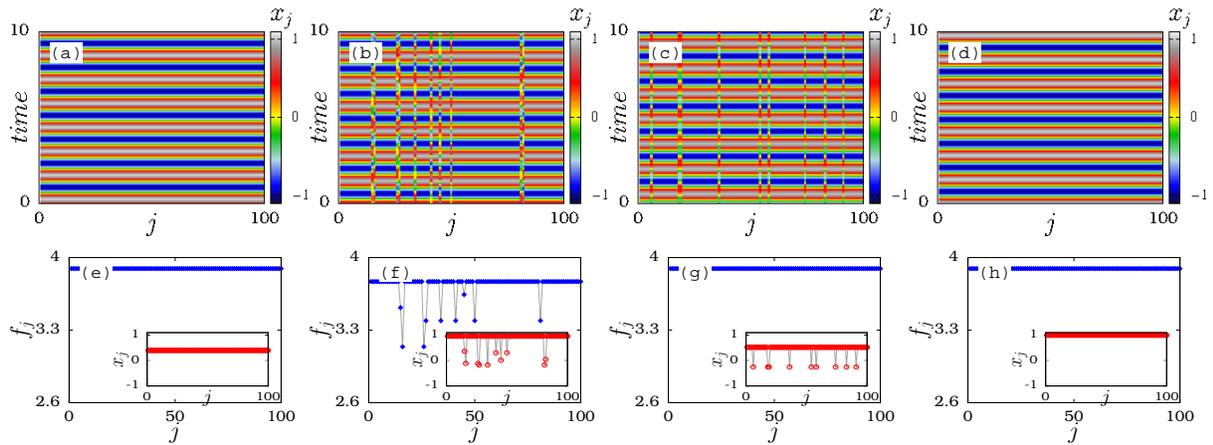}
		\caption{ Spatio-temporal plots for $q=0.1$ as a function of  $\epsilon$: (a) synchronized state for  $\epsilon=1.05$, (b) solitary state for $\epsilon=1.35$, (c) two cluster state for $\epsilon=1.65$ and (d) synchronized state for   $\epsilon=2.05$. The corresponding frequency plots are shown in the lower panels (e)-(h).  The insets in (e)-(h) denote the snapshots of the variables $x_i$.}
		\label{st} 
	\end{figure*} 
	\begin{figure*}[h]
		\centering
		%	\hspace{0.3cm}
		\includegraphics[width=17.0cm]{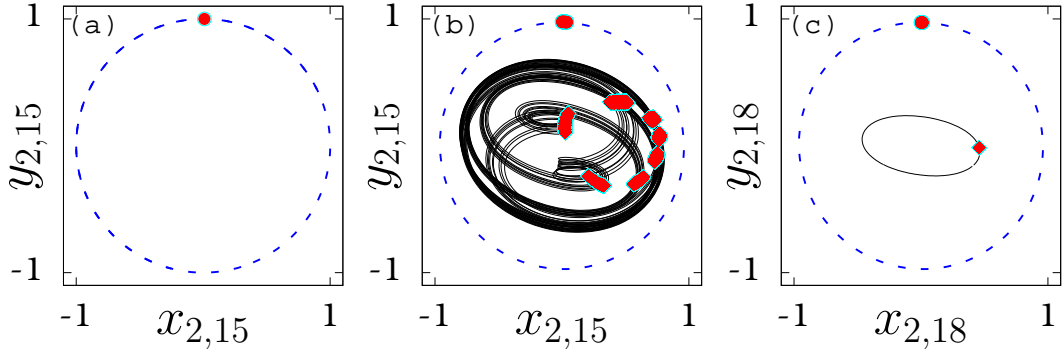}
		\caption{ Phase-portrait trajectories of the representative oscillators for (a)  synchronized state, (b) solitary state and (c) two-cluster oscillatory state. Dashed line denotes the phase-portrait trajectory of the synchronized group and solid line in Fig.~\ref{pp}(b) represents the  solitary oscillators,  while the solid line in Fig.~\ref{pp}(c) denotes second cluster.     Filled circles denote the Poincar\'e points corresponding to the synchronized group  whereas  Poincar\'e points for the quasiperiodic oscillations of the solitary oscillators are denoted  by filled diamonds.    The parameter values are  the same as in Figs.~\ref{st}(a)-(c).}
		\label{pp} 
	\end{figure*} 
	The emergence of swing by mechanism due to diffusive coupling  has been revealed both in the cases of  symmetry preserving  and symmetry breaking couplings, which induce  inhomogeneity among the synchronized regions resulting in the swing of synchronized states  \cite{main_prema}.  In order to understand  the impact  of the additional   repulsive coupling  on the emergence of the swing of the synchronized states in the ensemble of Stuart-Landau oscillators with  symmetry breaking coupling, at first, we have plotted the space-time evolution as a function of the coupling strength $\epsilon$ for  the  repulsive coupling $q=0.1$ (see Figs.~\ref{st}(a)-\ref{st}(d)).  Further to distinguish the distinct dynamical states, we have estimated the frequencies   of the oscillators corresponding to each of the observed  dynamical states,  which are depicted in Figs.~\ref{st}(e)-\ref{st}(h), while  the  snapshots of the dynamical states are shown in their respective insets.   From Figs. \ref{st}(a) and  \ref{st}(e), it is clear that a completely synchronized state (SYN) emerges for the coupling strength $\epsilon=1.05$.  In this state, all the oscillators oscillate with the same frequency as  is evident from Fig.~\ref{st}(e), while the inset in Fig.~\ref{st}(e)  clearly illustrates that each oscillator in the network oscillates with the  same amplitude.   Increasing the coupling strength  further to $\epsilon=1.35$, the system exhibits a solitary state (SS)   as shown in Figs.~\ref{st}(b) and  \ref{st}(f).  A very few of the oscillators makes an excursion away from the synchronized group and oscillate with their own frequency and amplitude  \cite{ss1,kapi1},  is evident from   Fig.~\ref{st}(f). In contrast to \cite{kapi1},  the solitary oscillators in Fig. 1(f) do not have identical mean frequencies.  In fact, their mean frequencies are distributed between four different levels.
%	{\bf In the ref. [51], the solitary state is characterized by all the solitary oscillators having the  same mean frequency.  Constrast to the earlier report,  the observed solitary oscillators  in Fig.~\ref{st}(f) oscillates with four different individual frequencies.} % The solitary state is characterized by same mean frequency of  all the solitary oscillators \cite{kapi1}. In contrast to the  ref.~\cite{kapi1},  the solitary oscillators  in Fig.~\ref{st}(f) oscillates with four different individual frequencies.} 
	Emergence of two-cluster (2C) state is observed on further increasing  the coupling strength to $\epsilon=1.65$ (see  Figs.~\ref{st}(c) and  \ref{st}(g)).  It is to be noted that in the 2C state,  the oscillators in the network oscillate in two different groups with different amplitudes but with the same frequency. 
	By increasing the coupling strength further, again the system re-enters into a  completely synchronized state (see  Figs.~\ref{st}(d) and  \ref{st}(h)), characterized by identical   amplitude and frequency.  Thus, it becomes evident that the phenomenon of swing of synchronized states, that is the  re-emergence of the synchronized state after being collapsed in a certain range of parameters, persists even with the addition of  repulsive coupling.
	For a clearer insight on the nature of the observed dynamical states, we have depicted the phase-space dynamics  of the  representative oscillators for  each of the observed states along with  their  Poincar\'e points  in Fig.~\ref{pp}.  The representative oscillator from the synchronized group is denoted by a dashed line, which always exhibits periodic oscillations.  The filled circles correspond to the Poincar\'e points of the synchronized group of oscillators.  The Poincar\'e points of the solitary oscillators are represented by  filled diamonds.  The representative oscillators from  the synchronized state $(x_2,x_{15})$ exhibiting periodic oscillations are evident from  Fig.~\ref{pp}(a) for $\epsilon=1.05$.  In the case  of the solitary state, the oscillator from the synchronized group $(x_{2})$  exhibits periodic oscillations as in Fig.~\ref{pp}(a), while the  solitary oscillator $(x_{15})$  exhibits quasi-periodic oscillations.   The closed loop of the  Poincar\'e points  confirms the quasi-periodic nature of the solitary oscillator.  The representative oscillators $(x_2,x_{18})$ from the two-cluster state (see Fig.~\ref{pp}(c)) indicate that the oscillators from the distinct clusters  follow different trajectories with different amplitudes.       
	
	Further to quantify the dynamical transitions and to characterize the swing of synchronized states, we have calculated  the standard deviation as a function of $\epsilon$ for three different values of the strength of the repulsive coupling.  The standard deviation is calculated  using the relation,
	\begin{eqnarray}
	\sigma =  \langle (\overline{|x_j-\overline{x_j}|^2})^{1/2} \rangle_t,
	\label{lst}
	\end{eqnarray}
	where the bar denotes  the  average over $1 \le j \le N$ and $< \cdot >_t$
	represents the time average.
	\begin{figure}[h]
		\centering
		\hspace{-0.4cm}
		\includegraphics[width=16.00cm]{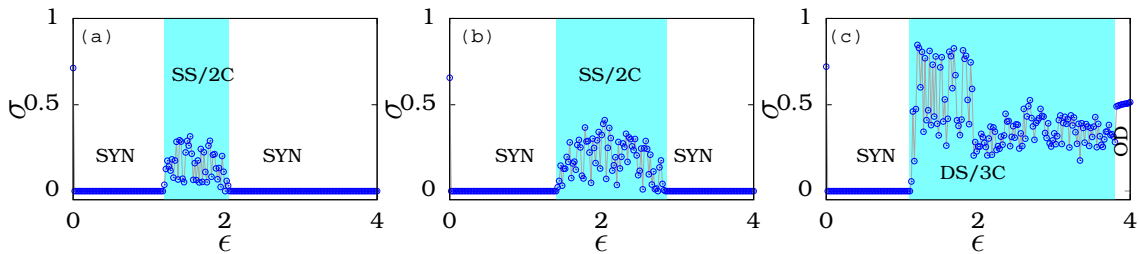}
		\caption{ Standard deviation as a function of $\epsilon$ (after leaving the transients time, $5\times10^4$ units) for (a) $q=0.1$ (b) $q=0.3$ and (b) $q=0.7$. SYN, SS and DS denote the synchronized state, solitary and desynchronized states.  2C and 3C are the two and three cluster oscillatory states.  OD is the oscillation death state.}   
		\label{stand} 
	\end{figure} 
	Initially, the dynamical transitions will be analyzed for a smaller value of the strength of the   repulsive coupling. The standard deviation is shown in Fig.~\ref{stand}  as a function of the coupling strength $\epsilon$ for three different values of the repulsive coupling. Fig.~3(a) corresponds to the strength of the repulsive coupling $q=0.1$. It is evident from the null value of the standard deviation in the range of $\epsilon\in(0.08,1.09)$ that  all the oscillators in the network evolve in synchrony with each other. Further, the finite value of the standard deviation in a narrow range of $\epsilon\in(1.09,1.78)$ indicates that  all the oscillators in the network  evolve independently (asynchronously).  The oscillators in this range of $\epsilon$  actually exhibit either solitary state or two cluster state.  The null value of the standard deviation beyond $\epsilon=1.78$ elucidates that all the oscillators are in  complete synchrony.  Thus the value of the standard deviation as a function of the coupling strength $\epsilon$  corroborates the re-emergence of the synchronized state after a short range of  excursion away from the synchronized state,  thereby illustrating the phenomenon of the swing of the synchronized state. 
	
	\par Similarly, Figs.~\ref{stand}(b)  and \ref{stand}(c)  are plotted for two distinct values of the strength of the repulsive coupling (i.e. $q=0.3$ and $q=0.7$).  It is clear that increasing $q$ increases the range of disorder among the coupled oscillators as a function of the coupling strength $\epsilon$ (see Figs.~\ref{stand}(a)-\ref{stand}(c)). The synchronized state re-emerges for $q=0.3$ as a function of the coupling strength $\epsilon$ as indicated by the value of the standard deviation in Fig. 3(b). It is also to be noted that the synchronized state characterized by the null value of the standard deviation does not re-emerge for a larger value of the repulsive coupling strength (see Fig. 3(b) for $q=0.7$) thereby elucidating the disappearance of the swing by mechanism for larger ranges of the strength of the repulsive coupling. In Figs.~\ref{stand}(a)-\ref{stand}(c), the shaded regions denote the desynchronized (DS) or three-cluster (3C) (disordered) states, whose spread  increases with   the strength of  the repulsive coupling.   Also, we observed the oscillation death state at strong coupling strengths (see Fig.\ref{stand}(c)) which will be detailed in the following.
	\begin{figure*}[h]
		\centering
		\hspace{-0.1cm}
		\includegraphics[width=12.0cm]{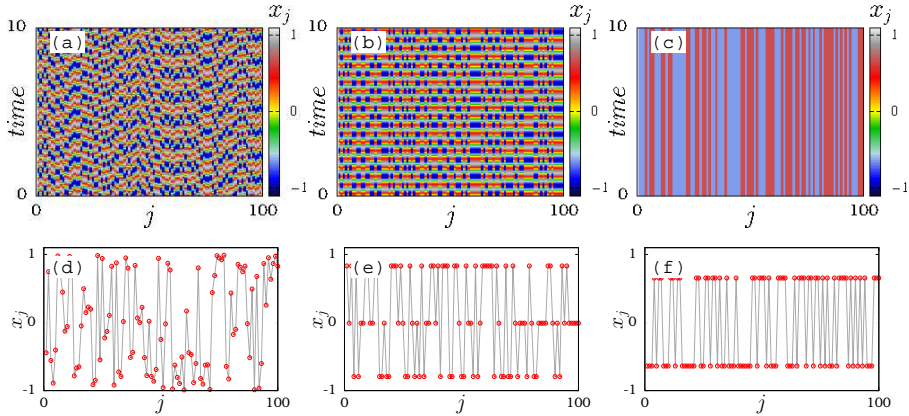}
		\caption{ Spatio-temporal evolutions for $q=0.7$, as a function $\epsilon$, (a) desynchronized  (DS) state for  $\epsilon=1.3$ (b) three  cluster (3C) state for $\epsilon=2.8$, (c)  oscillation death state for $\epsilon=4.0$. The corresponding snapshots of the variables $x_i$ are shown in the lower panels (d)-(f).  }
		\label{st_q1} 
	\end{figure*} 
	\par	To understand the dynamical transitions at even stronger repulsive coupling strengths,  we have plotted space-time plots  and snapshots for $q=0.7$ for three different values of  $\epsilon$ in Fig.~\ref{st_q1}. From the figure, it is evident that there is a transition from desynchronized state to oscillation death state through the three-cluster state as a function of the  coupling strength $\epsilon$ for the chosen value of the strength of  the repulsive coupling.  The coupled Stuart-Landau oscillators evolve independently  as shown in Figs.~\ref{st_q1}(a) and \ref{st_q1}(d) for $\varepsilon=1.3$.  The asynchronous oscillators get segregated into three clusters for further increase in the attractive coupling as depicted in Figs. \ref{st_q1}(b) and \ref{st_q1}(e) for $\epsilon=2.8$. Finally, the oscillators randomly populate the upper and lower branches of the inhomogeneous steady state as shown in Figs.~\ref{st_q1}(c) and \ref{st_q1}(f) for $\epsilon=4.0$. The global dynamical transitions will be delineated in the following. 
	
\subsection{Global dynamical behavior of coupled oscillators}
	\begin{figure}[h]
		\centering
		\hspace{-0.1cm}
		\includegraphics[width=10.0cm]{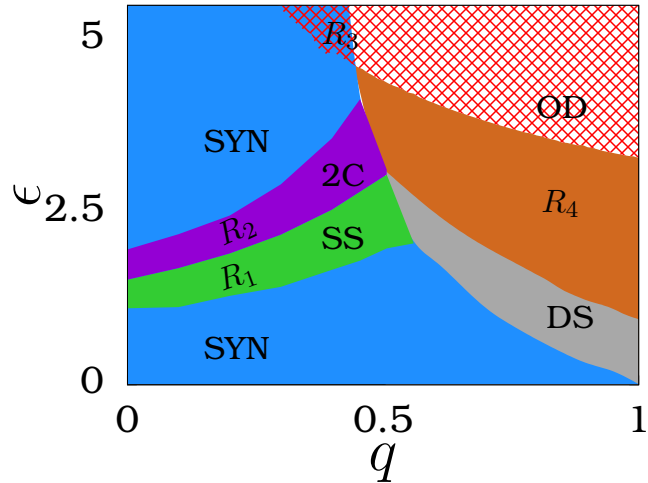}
		\caption{ Two parameter phase diagram  for globally coupled Stuart-Landau oscillators in  $(q,\epsilon)$ space.   SYN, SS and DS are the synchronized, solitary and desynchronized states, respectively.  2C and 3C denote the two-cluster and three-cluster oscillatory states.  The oscillation death is denoted by OD.  $R_1$, $R_2$, $R_3$ and $R_4$ are the multistability regions between SYN-SS, SYN-2C, SYN-OD and DS-3C, respectively.}
		\label{glob} 
	\end{figure}
	\par 	To unravel the global dynamical behavior of the  coupled Stuart-Landau oscillators, we have depicted a two parameter phase diagram in the  $(q,\epsilon)$ space in Fig.~\ref{glob}.   The coupled Stuart-Landau oscillators  evolve in synchrony for lower values of  attractive  $(\epsilon)$ and repulsive   $(q)$ couplings.  For  lower values of the strength of repulsive coupling ($q<0.5$), the dynamical system admits the  re-emergence of synchronized state via the solitary and two-cluster states as a function of the coupling strength, thereby corroborating the existence of the phenomenon of the swing of the synchronization in a rather large range of the parameters.  Further, there is a coexistence of the synchronized state with the solitary state and two-cluster state in the regimes $R_1$ and $R_2$, respectively.   For a larger value of $\epsilon$ (approximately around $\epsilon=4.5$), and $q$ around $0.45$, there is a  coexistence of the oscillation death state (indicated by the checked regimes) with the synchronized state in the region $R_3$.  The complete synchronization  region is obtained by choosing  nearly  the same initial states for all the oscillators. For larger values of the repulsive coupling strength, the coupled system exhibits transition from homogeneous oscillatory state (synchronized state) to  inhomogeneous steady state (oscillation death state) via asynchronous and three-cluster states as a function of  $\epsilon$.  In the oscillation death state the oscillators populate both the upper and lower branches of the inhomogeneous steady state randomly. In addition, there is a coexistence of desynchronized state with the three-cluster state in the region denoted by $R_4$. Thus it is evident that smaller values of the repulsive coupling strength favors the  onset of the phenomenon of the swing of the synchronized state.  On the other hand, stronger repulsive coupling favors the independent evolution of the oscillators thereby preventing them from entrainment with each other in  a large range of the attractive coupling $\epsilon$ as is evident from Fig.~\ref{glob}.  In addition, the strong repulsive coupling favors the  onset of the inhomogeneous death state in a large range of $\epsilon$.
	
	The numerical boundaries of each regions were distinguished  by using the notions of  $D$-factor  \cite{si1} and  the strength of incoherence $(SI)$ \cite{main_prema, si2}.  Initially, the oscillatory and oscillation  death state are identified through the $D$-factor.  Then the distinct oscillatory dynamical behaviors are classified from the $SI$,   which takes  the values $0$ and $1$ for  synchronized  and desynchronized states, respectively.  For the two-cluster  (2C), three cluster (3C) and solitary state (SS),  the values of $SI$ lies between $0< SI < 1$.  Further,  2C,  3C  and SS states are distinguished by using the removal of  discontinuities  \cite{si2}. The analytical stability conditions for the observed dynamical states are detailed in the next section. 
	\section{Analytical results}\label{anal}
	In the following, we will deduce the analytical stability conditions for the observed dynamical states.
	\subsection{Stability conditions for distinct oscillatory states}
	To start with, we consider the solitary state which is characterized by a very few oscillators making excursion away from the majority of the synchronized oscillators.  Hence, the equation corresponding to the coupled Stuart-Landau oscillators system (1), can be  decomposed as, 
	\begin{eqnarray}
	\fl 	\frac{dz_{s}}{dt} = f(z_{s})+\epsilon[p Re(Z_{s}) + \bar{p} Re(Z_{d})-Re(z_{s})]
	-iq\epsilon[p Im(Z_{s}) + \bar{p} Im(Z_{d})-Im(z_{s})] ,  \\
	\label{syn}
	\fl	\frac{dz_{d}}{dt} = f(z_{d})+\epsilon[p Re(Z_{s}) + \bar{p}Re(Z_{d})-Re(z_{d})] 
	-iq\epsilon[p Im(Z_{s}) + \bar{p} Im(Z_{d})-Im(z_{d})] ,
	\label{ds_eq4}
	\end{eqnarray}
	where $z_s$  represents the state of synchronized oscillators and $z_d$  represents the desynchronized (solitary)  oscillators. Further, we have $f(z_{s}) =(1+i\omega)z_{s}-(1-ic)|z_{s}|^2 z_{s}$  and $f(z_{d})=(1+i\omega)z_{d}-(1-ic)|z_{d}|^2 z_{d}$, $\bar{p} = 1-p$,  $s=1,2,...l_1$, $d=1,2,...l_2$,  $l_1+l_2=N$. Here  $N$ is the total number of oscillators in the network,  $p=l_1/N$, $\bar{p}=l_2/N$, $p+\bar{p}=1$.  Also $Z_s=\frac{1}{l_1}\sum_{s=1}^{l_1} z_s$ and $Z_d=\frac{1}{l_2}\sum_{d=1}^{l_2} z_d$.  
	
	If we choose $p\approx 1$, then $\bar{p}\approx0$ and hence the synchronized oscillators are decoupled from the solitary oscillators (see Eq.~(3) and Eq.~(\ref{ds_eq4})) \cite{daido_anal1, daido_anal2, main_prema, vkc_anal,gopal_anal}.  As a consequence,  the solution of the  synchronization state  can be deduced as $z_s=e^{i(\omega+c)t}$.  Now, upon substituting $z_s$,   Eq.~(\ref{ds_eq4}) can be rewritten as 	
	\begin{equation}
	\frac{dz_{d}}{dt} = f(z_{d})+\epsilon[Re(e^{i(\omega+c)t})-Re(z_{d})] 
	-iq\epsilon[Im(e^{i(\omega+c)t}) -Im(z_{d})] .
	\label{mts}
	\end{equation}
	The above equation corresponding to the solitary oscillator is also found to admit the solution of  the synchronized state  $z_d=z_s=e^{i(\omega+c)t}$, which may be stable or unstable depending on the strength of the repulsive coupling. Thus it is evident that the above synchronization manifold is one of the possible dynamical states of the coupled Stuart-Landau oscillators. 
	
	In order to unravel the other possible collective dynamical behaviors, we apply  the ansatz  $z_d(t)= Z(t)e^{i(\omega+c)t}$, where
	$Z(t)$ is the amplitude of the desynchronized state. Inserting this into  Eq.~(\ref{mts}), we obtain	          
	\begin{eqnarray}
	\frac{dZ}{d\tau} = (1-ic) Z-(1-ic)|Z|^2 Z+ \frac{1}{2}(1-Z)\epsilon \bar{q}+ \frac{1}{2}(1-\bar{Z})\epsilon \bar{q} e^{-2i(\omega+c)t},
	\label{pert}
	\end{eqnarray}
	where $\bar{q}=1-q$.  Assuming that $\omega +c >>1$, we can averge Eq.~(\ref{pert}) over the fast oscillation period $[0,2\pi/(\omega+c)]$. This yields
	\begin{eqnarray}
	\frac{dZ}{d\tau} = (1-ic) Z-(1-ic)|Z|^2 Z+ \frac{1}{2}(1-Z)\epsilon \bar{q}.
	\label{per_eq}
	\end{eqnarray} 	
	Now, Eq.~(\ref{per_eq}) has the following three  fixed points
	%\begin{subequations}
		\begin{eqnarray}
		Z_1=1, \\
		\label{fp_1}
		|Z_{2,3}|^2= \frac{(1+c^2)-\epsilon\bar{q}\mp \sqrt{(1-c^2-\epsilon\bar{q})^2-(1+c^2)\epsilon^2\bar{q}^2}}{2(1+c^2)}. 
		\label{fp_2}
		\end{eqnarray} 
	%\end{subequations}
The  stability of the fixed point $Z_1=1$ corroborates the stability of the   complete synchronization manifold  and  the other two fixed points determine the stability of desynchronized (solitary) oscillator which exist when,   
	\begin{eqnarray}
	\epsilon<\epsilon_{SN}  = \frac{1-c^2(q-1)\sqrt{(1+c^2)^3(q-1)^2}+q}{(1+c^2) (q-1)^2}.
	\end{eqnarray} 
	Carrying out the linear stability analysis  of Eq.~(\ref{per_eq}), we find that the determinant of the Jacobian matrix for the fixed points $Z_2$ and $Z_3$ can be expressed as
	\begin{equation}
	det (J_Z{_{_{2,3}}}) = \frac{c^6-1-d_1c^4+d_2c^2-2\epsilon(q-1) \mp \sqrt{\Delta} d_3}{2(c^4-1)},
	\label{det}
	\end{equation}
	where $d_1=\epsilon^2(q-1)^2-1-2\epsilon(q-1)$, $d_2=\epsilon^2(q-1)^2-1$, $d_3=1+c^2-\epsilon$ and $\Delta= (c^2-1)^2 (1+c^4-c^2(\epsilon^2(q-1)^2-2(1+\epsilon)(q-1))+2\epsilon(q-1))$. The value of the determinant  is negative  for $Z_2$, and positive for $Z_3$, which imply  that the fixed point $Z_2$ turns out to be  a saddle whereas the other fixed point  $Z_3$ is either an unstable node or a focus for $|c| \le 1$.  Further the unstable node or focus turns out to be a stable attractor when $|c| > 1$ and  $\epsilon>\epsilon_{HB}$, where $\epsilon_{HB}$    is given by

	\begin{figure}[h]
		\centering
		\hspace{-0.1cm}
		\includegraphics[width=14.5cm]{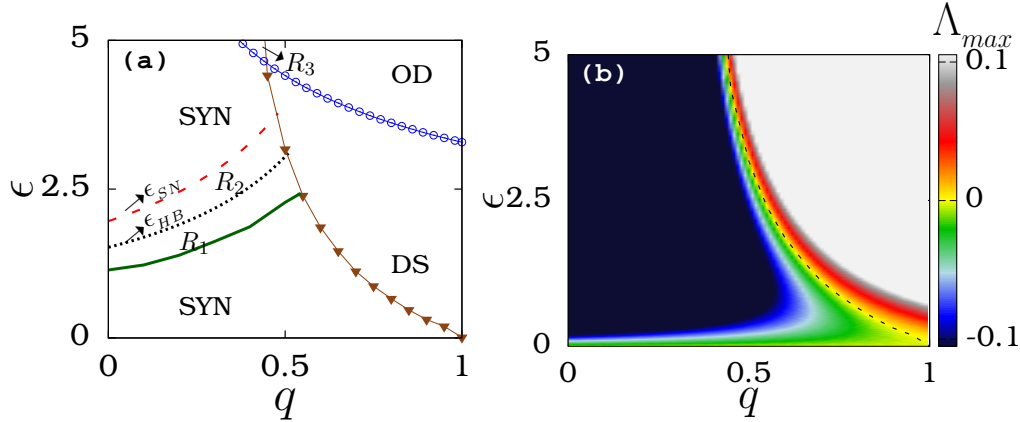}
		\caption{ (a) Analytical stability curves for distinct dynamical states in $(q, \epsilon)$ space. Dashed and dotted curves represent the saddle node and Hopf bifurcation curves. The solid  curve is the numerically obtained boundary.  The lines connected by filled triangles and open circles denote the boundary for synchronized and oscillation death states, respectively.  $R_1$, $R_2$ and $R_3$  are the multistability regions between SYN-SS, SYN-2C and SYN-OD, respectively.  (b) The largest Lyapunov exponent $\Lambda_{max}$  in the  $(q, \epsilon)$ space. The dotted line in Fig.~(b) denotes the stable boundary of the synchronized state.}		
\label{bound} 
	\end{figure} 
		\begin{eqnarray}
	\epsilon_{HB}  = \frac{2(2-\sqrt{(5+2c^2-c^4)(q-1)^2}-2q)}{(1+c^2) (q-1)^2},
	\end{eqnarray} 
	which is obtained by equating the trace  of the  Jacobian matrix of Eq.~(\ref{per_eq}) to zero $(tr(J_Z{_{_{3}}})=0)$, so that
	\begin{eqnarray}
	tr(J_Z{_{_{3}}})= \frac{1}{(c^4-1)}[2\Delta+(c^2-1)^2\epsilon(q-1)].
	\end{eqnarray}
	
	The Hopf bifurcation and saddle node  curves are denoted by dotted  and dashed lines, respectively, in Fig.~\ref{bound}(a).   The lower boundary, indicated by the solid (green) line, separating the completely synchronized state and the multistability state $R_1$ is obtained by solving Eq.~(\ref{per_eq}) numerically.   In the  region $R_1$, the system exhibits solitary state, where a small number of  individual oscillators excurs away from the synchronized group  and oscillates with quasi-periodic oscillations. The solitary state further gets destabilized via $\epsilon_{HB}$ and gives birth to two-cluster (2C)  oscillatory state, which is denoted by  the region $R_2$.  In the 2C state, the oscillators  in the network split  into two groups  and oscillate with different amplitudes. On the other hand   the cluster state  (see region $R_2$) is stabilized via $\epsilon_{HB}$ which is further destabilized via   $\epsilon_{SN}$ (dashed line). 
		\par In addition,   from the linear stability criterion for the solution $Z=1$ given   in Eq.~(8) one can show that the  synchronized state is stable in the entire region of $(q,\epsilon)$ space.  But it is known that the  repulsive coupling can destabilize the synchronized region at strong repulsive coupling strengths in the parametric space.   In order to find the   analytical condition for the  synchronized manifold and to find the limit of stable solitary and two-cluster states as a function of the repulsive coupling strength,  we use  the master stability approach	\cite{msf1}-\cite{msf3}.   In the synchronized manifold,  $x_k=x$,  and $y_k=y$, $\forall\, k=1,2...,N$, where $x(t)$ and $y(t)$ are the solutions for the uncoupled  system  (\ref{model1}).   For an infinitesimal perturbation to the synchronized solution,  Eq.~(\ref{model1})   can be reduced as
	%The variational equations for the perturbations  { $ \beta_{jk}$} to the synchronized manifold   is defined as 
	\begin{eqnarray}
	\fl	{\dot{\boldsymbol \eta_1}} = (1-3x^2-y^2-2cxy){\boldsymbol \eta_1}-(\omega+cx^2+3cy^2+2xy){\boldsymbol \eta_2}+\epsilon {\bf G} {\boldsymbol \eta_{1}}, \nonumber \\
	\label{var1}
	\fl	{\dot{ \boldsymbol {\eta_{2}}}}= (\omega+3cx^2+cy^2-2xy){\boldsymbol \eta_1}+(1-x^2-3y^2+2cxy){\boldsymbol \eta_2}-q\epsilon {\bf G}  {\boldsymbol \eta_{2}} ,\\ 
	\qquad 	\qquad 	\qquad 	\qquad 	\qquad 	\qquad 	\qquad 	\qquad 	\qquad 	\qquad 	\qquad {k=1,2,...N} \nonumber
	\end{eqnarray} 
	%Where, ${\bf 1}_N$ is the N-dimensional unitary matrix, {\bf  F(z)} is the uncoupled system and ${D\bf H(z)}$ coupling connectivity matrix.  By diagonalizing  each block in  Eq.~(\ref{v1}), the first block is not affected by the transformation and the second block only diagonalizing the matrix $\bf G$. Then the variational equation having the form, 
	%	
	Here $\boldsymbol{{\eta_1}}=(\eta_1^x,\eta_2^x,...,\eta_N^x)^T$ and $\boldsymbol{{\eta_2}}=(\eta_1^y,\eta_2^y,...,\eta_N^y)^T$. We note that {$\eta_j^x$} and {$\eta_j^y$} are the deviations of $x_j$ and $y_j$ from the  synchronized solution $(x, y)$.  ${\bf G}$ is  the coupling matrix which %  corresponding to the eigenvalue $\lambda_k$. The coupling matrix
	is defined as
 \[
   {\mathbf{G}}= \frac{1}{N}
  \left( {\begin{array}{cccc}
   1-N & 1 & \cdots & 1 \\
   1 & 1-N & \cdots & 1 \\
   \vdots & \vdots &   \ddots  & \vdots \\
 {1} & {1} &    \cdots & {1-N} 
  \end{array} } \right)
\]
Then  the perturbation from the synchronized manifold is defined as	  ${\beta_{jk}}$ $(j=1,2)$, where   $\beta_{jk}={\boldsymbol {\eta_j^T V_k}}$.  $\boldsymbol{V_k}$ is the  eigenvector of the coupling matrix {\bf G}. The diagonalized variational equation from Eq.~(\ref{var1}) can be written as, 
	\begin{eqnarray}
	\fl	{\dot{\beta}_{1k}} = (1-3x^2-y^2-2cxy)\beta_{1k}-(\omega+cx^2+3cy^2+2xy)\beta_{2k}+\epsilon \lambda_k\beta_{1k}, \nonumber \\
	\label{msf_eq}
	\fl	{\dot{\beta}_{2k}}= (\omega+3cx^2+cy^2-2xy)\beta_{1k}+(1-x^2-3y^2+2cxy)\beta_{2k}-q\epsilon \lambda_k\beta_{2k},\\ 
	\qquad 	\qquad 	\qquad 	\qquad 	\qquad 	\qquad 	\qquad 	\qquad 	\qquad 	\qquad 	\qquad {k=1,2,...N} \nonumber
	\end{eqnarray}  
	
	where $\lambda_k'$s are the eigenvalues of the coupling matrix {\bf G}. The stable region of the synchronized state is obtained from  the eigenvalues of the coupling matrix {\bf G},  which	turn out to be $\lambda_0=0$, $\lambda_k=-1$, $k=1,2,...N-1$.  The eigenvalue  $\lambda_0=0$ corresponds to  the perturbation parallel to the synchronization manifold, while the other eigenvalues correspond to the perturbations  transverse to the synchronization manifold. The transverse eigenmodes should be damped out to have a stable synchronization manifold.  By substituting the eigenvalues  in  Eq.~(\ref{msf_eq}) and by finding the largest  Lyapunov  exponents ($\Lambda_{max}$), one can obtain the stable boundaries of the synchronized region  which separates the boundary between synchronized and desynchronized states.  Whenever the largest  Lyapunov  exponents acquires negative value the   synchronized manifold is stable.    The line connecting the filled triangles     in Fig.~\ref{bound} (a) denote the  boundary between the synchronized and desynchronized states and the corresponding largest Lyapunov exponents in  $(q, \epsilon)$ space is shown in Fig.~\ref{bound} (b).
	Finally, we will deduce the analytical stability condition for oscillation death state in the following. 
	\subsection{Stability condition for oscillation death state}
	The system of  coupled  Stuart-Landau oscillators also exhibits oscillation death state at strong coupling strengths, where the system suppresses the oscillations and ultimately attains stable inhomogeneous steady states.   Here, we observed the total population of the network splits into  either upper or lower branches of inhomogeneous states.  The corresponding dynamical equations can be written as, 
	\begin{eqnarray}
	\frac{dz_{d1}}{dt} = f(z_{d1})+\epsilon[p Re(z_{d1}) + \bar{p}  \nonumber Re(z_{d2})-Re(z_{d1})] \\ 
	\qquad \qquad \qquad \qquad \quad \quad \quad \quad	-iq\epsilon[p Im(z_{d1}) + \bar{p} Im(z_{d2})-Im(z_{d1})], 
	\label{f1}
	\end{eqnarray}
	\begin{eqnarray}		
	\frac{dz_{d2}}{dt} = f(z_{d2})+\epsilon[p Re(z_{d1}) + \bar{p} Re(z_{d2})-Re(z_{d2})]  \nonumber \\
	\qquad \qquad \qquad \qquad \quad \quad \quad \quad	-iq\epsilon[p Im(z_{d1}) + \bar{p} Im(z_{d2})-Im(z_{d1})],
	\label{f2}
	\end{eqnarray}
	where $f(z_{d1})=(1+i\omega)z_{d1}-(1-ic)|z_{d1}|^2 z_{d1}$ and $f(z_{d2})=(1+i\omega)z_{d2}-(1-ic)|z_{d2}|^2 z_{d2}$.  The fixed point solutions for $p=0.5$, of Eq.~(\ref{f1}) and Eq.~(\ref{f2}) can be expressed as	
	\begin{eqnarray}
	x_{1,2}^* &=& \mp \frac{((1+q)\epsilon-f_0)}{\sqrt{2} c(\epsilon-1)-\omega}  \sqrt{\frac{1}{(1+c^2)^2(1+q)\epsilon}(f_1\epsilon^2+f_2\epsilon+f_3),} \nonumber \\ 
	y_{1,2}^* &=& \frac{c(\epsilon-1)-\omega}{((1+q)\epsilon-d)} \,\,x_{1,2}^*\,,
	\end{eqnarray}
	where $f_0= \sqrt{h_1\epsilon^2+h_2(1-q)\epsilon-4h_3}$, $f_1=q(1+q)+c^2(3q-1)$, $f_2=(c^2+q)f_0+1+5c^2+q(1-3c^2)+2c\omega(1-c^2)-4cq\omega$, $f_3=h_4f_0+2c^3\omega-4c^2-6c\omega-2\omega^2(1-c^2)$, 
	$h_1=q^2+4c^2q+2q+1$, $h_2=4c(c+\omega)$, $h_3=c^2+2c\omega+\omega^2$ and $h_4=1-c^2-2c\omega$.  In the oscillation death state,   each oscillator in the network  settles down in one of the inhomogeneous steady states. Further using the linear stability analysis, we can find the stability conditions for oscillation death state.
	The  inhomogeneous steady state becomes stable at 
	\begin{equation}
	\epsilon= \frac{1}{4q}(q-1) +\sqrt{1+q(2+4c^2+16\omega(c+\omega)+q)}.
	\end{equation}
	The stable boundary (line connected by unfilled circles) of OD  region is shown in Fig. \ref{bound}. The stable boundary of OD state agrees well  with the numerical boundary.  In addition,  there is a  coexistence of OD with SYN state in the region $R_3$ (see Fig.~\ref{bound}).    From the stable boundaries of the SYN, SS and 2C states,  the swing of synchronized state is strongly evident.    The obtained stable boundaries of the  synchronized state, solitary and cluster states in Fig.~\ref{bound} are  in good agreement with the numerical  boundaries between the dynamical states in Fig.~\ref{glob}.  
	\section{\label{plc} Dynamical transitions for varying power-law exponent $(\alpha \ge 0)$}
	\par Previously, we have analysed the dynamical transitions for power-law exponent $\alpha=0$.  In this section, we study the dynamical behavior %while changing the network from global coupling topology  to local
	while varying the power-law exponent $(\alpha)$  \cite{tbaner1,tbaner2}, by considering the ring network of coupled Stuart-Landau oscillators (1) for  $N=101$.  
	
	\begin{figure}[h]
		\centering
		\hspace{-0.4cm}
		\includegraphics[width=16.00cm]{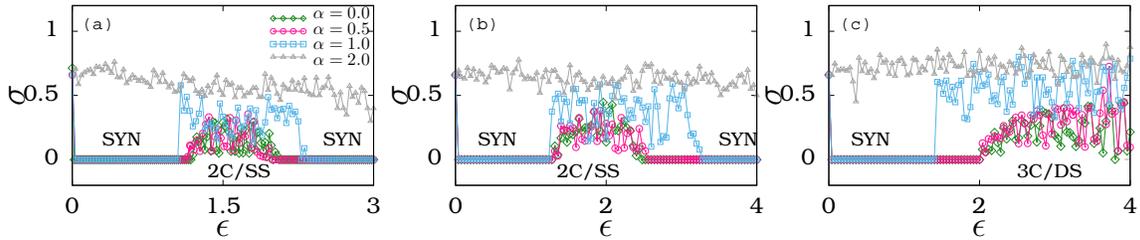}
		\caption{ Standard deviation as a function of coupling strength  $\epsilon$  (after leaving the transients time, $5\times10^4$ units) for (a) $q=0.1$ (b) $q=0.3$ and (b) $q=0.7$. The open diamonds, circles, squars and triangular points connecting by lines denote the power-law exponent $\alpha=0.0,~\alpha=0.5,~\alpha=1.0$ and $\alpha=2.0$, respectively. }
		\label{stand_s}
	\end{figure}
	Figs.~\ref{stand_s}(a)-\ref{stand_s}(c) are plotted for three  different values of the repulsive coupling strengths $q=0.1$, $q=0.3$ and $q=0.7$, as a function of coupling strength  $\epsilon$. The different lines connected by points, namely diamonds, circles, squares and triangles, respectively represent  the power-law exponents $\alpha=0.0,~\alpha=0.5,~\alpha=1.0$ and $\alpha=2.0$.  Fig.~\ref{stand_s}(a) clearly illustrates the emergence of  the  swing of synchronized state, for $\alpha=0$. Increasing  $\alpha$ to  0.5 and 1.0 give rise to  a widening of the  disorder region among the synchronized region. Upon increasing $\alpha$ to still higher values $(\alpha=2.0)$, the system exhibits  completely desynchronized (disordered) state for the entire range of coupling strength  $\epsilon$. Similar dynamical transitions are observed for $q=0.3$ and $q=0.7$ in Figs. \ref{stand_s}(b) and \ref{stand_s}(c).   From Figs.~\ref{stand_s}, we have identified the fact that  increasing the power-law exponent can increase the disorderliness among the dynamical regions. 
	\begin{figure}[h]
		\centering
		\hspace{-0.4cm}
		\includegraphics[width=10.00cm]{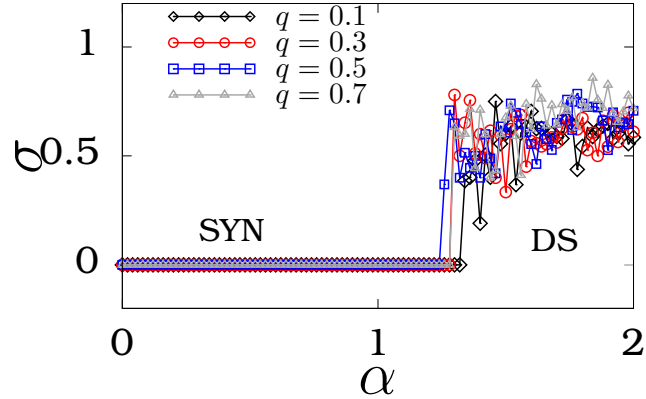}
		\caption{ Standard deviation as a function $\alpha$ by fixing     $\epsilon=1.0$  for different values of $q$  (after leaving the transients time, $5\times10^4$ units).  The open diamonds, circles, squares and triangular points connected by lines denote the power-law exponents for $q=0.1,~q=0.3,~q=0.5$ and $q=0.7$, respectively. }
		\label{s_vary}
	\end{figure} 
	In order to show this  clearly we have also plotted the  dynamical transitions  as a function of power-law exponent by fixing the coupling strength at $\epsilon=1.0$  for four different strengths of repulsive coupling $q$ (see Fig.~\ref{s_vary}).  The different lines connected by points such as diamonds, circles, squares and triangles represent  different values of the strength of  repulsive coupling, namely $q=0.1,~q=0.3,~q=0.5$ and $q=0.7$, respectively. For all the values of $q$, the system is synchronized upto a critical value of $\alpha$ (i.e. $\alpha \approx 1.2$), which  disappears  while increasing the  power-law exponent $\alpha$ to further  higher values  and  the system becomes completely desynchronized (disordered).
	\section{Conclusion}\label{con} 
	Long-range interacting systems are very rich and intriguing due to their omnipresent nature  in real world applications. Recently, many investigations have been carried out for an in-depth understanding of long-range interacting systems, including galaxies, quantum spin models  and cold atom models.  In this article, we have investigated the collective dynamical behavior in  a long range interacting system of coupled Stuart-Landau limit cycle  oscillators with a combination of attractive and repulsive couplings as a prototypical example. The dynamical behaviors were found as a function of the  strength of  the repulsive coupling.  Interestingly, we find  the emergence of the  swing of the synchronized state,  where the stable synchronized  state is destabilized by the emergence of  solitary or cluster  states  which is again   stabilized while increasing the coupling strength.  This re-emergence of synchronized state is called  the swing by mechanism.   We  have also found  increasing   disorder among the coupled Stuart-Landau oscillators   when increasing the strength of the repulsive coupling.  Finally, we have deduced the analytical boundary condition for the   synchronized state, solitary and cluster states and  also for oscillation death state. We have  also found that the analytically obtained boundaries  match perfectly  with the numerical boundaries.  The dynamical transition is also  observed as a function of power-law exponent and we find that the system exhibits completely desynchronized (disordered) state while increasing the power-law exponent.   The observed results  may  shed light for the control of complete synchronization in long-range  interacting systems. 
	\section*{Acknowledgments}
	KS sincerely thanks the CSIR for a fellowship under SRF Scheme (09/1095(0037)/18-EMR-I).  The work of VKC forms part of a research project  SERB-DST Fast Track scheme for young scientists under Grant No. YSS/2014/000175 and the CSIR Project under Grant No.03(14444)/18/EMR II. DVS is supported by the CSIR EMR Grant No. 03(1400)/17/EMR-II. The work of ML is supported by DST-SERB Distinguished Fellowship.  
	%\end{acknowledgments}


\section*{References}
	\begin{thebibliography}{99}	
		\bibitem{glo1} De Monte S, d'Ovidio F  and Mosekilde E 2003  Coherent regimes of globally coupled dynamical systems \textit{Phys. Rev. Lett.} {\bf{ 90}}, 5
		
		\bibitem{glo2}  Sailer X,  Zaks M, and Schimansky-Geier L 2005 Collective dynamics in an ensemble of globally coupled FHN systems \textit{Fluctuation and Noise letters} {\bf 5} L299-L304
		
		\bibitem{appl_1}  Moon J-Y, l Lee U,  Blain-Moraes S,  Mashour G A 2015 General relationship of global topology, local dynamics, and directionality in large-scale brain networks \textit{PLoS Comput. Biol.} {\bf{11}} e1004225
		
		\bibitem{rev2_glob}   Pikovsky A and  Rosenblum M 2015 Dynamics of globally coupled oscillators: Progress and perspectives  \textit{Chaos} {\bf{25}} 097616
		
		\bibitem{chemic1}   Kuramoto Y  1981 Rhythms and turbulence in populations of chemical oscillators   \textit{Physica A: Statistical Mechanics and its Applications} {\bf{106}} 128-143.
		
		\bibitem{chimec3}    Kuramoto Y, \textit{Chemical  Oscillations, Waves  and  Turbulence} (Springer, Berlin, 1984).
		
		\bibitem{phys1}  Chavanis P H and  Rieutord M 2003 Statistical mechanics and phase diagrams of rotating self-gravitating fermions \textit{Astronomy and Astrophysics} {412} 1. 
		\bibitem{phys2}  Vega H J and  S\'anchez 2002 Statistical mechanics of the self-gravitating gas: I. Thermodynamic limit and phase diagrams  \textit{Nuclear Physics B} {\bf 625} 407-518.
		
		\bibitem{phys3}  Bouchet F and Sommeria J 2002 Emergence of intense jets and Jupiter's great red spot as maximum-entropy structures \textit{Journal of Fluid Mechanics} {\bf 464} 165-207.
		
		\bibitem{phys4}  Bouchet F and Barr\'e J   Statistical mechanics of systems with long range interactions  \textit{J. phys. conf. series} {\bf 31} 18-26.  % main paper physical systems
		
		\bibitem{elect}   Hadley P,  Beasley M R, and  Wiesenfeld K  1988 Phase locking of Josephson-junction series arrays \textit{Phys. Rev. B} {\bf 38} 8712
		
		\bibitem{bio}   Glass L  2001  Synchronization and rhythmic processes in physiology \textit{Nature (London)} {\bf 410} 277;  Mosekilde E,
		Maistrenko Y, and  Postnov D 2002 Chaotic Synchronization: Applications to Living Systems  \textit{Chaotic Synchronization} (World    Scientific,    Singapore, 2002); 
		
		
		
		\bibitem{bio1} Lerch H P,    Stange P,      Mikhailov A S,   and    Hess  B 2002 Mutual synchronization of molecular turnover cycles in allosteric enzymes III.  Intramolecular Cooperativity \textit{J.   Phys. Chem. B} {\bf 106} 3237
		
		\bibitem{scho_glob}  B\"ohm F, Zakharova A, Sch\"oll E, L\"udge K  2015 Amplitude-phase coupling drives chimera states in globally coupled laser networks \textit{Phys. Rev. E} {\bf 91} 040901
		
		\bibitem{clust1} Schmidt L and  Krischer K 2015 Clustering as a prerequisite for chimera states in globally coupled systems
		\textit{Phys. Rev. Lett.} {\bf{ 114}} 034101
		
		\bibitem{clust2} Schmidt L and  Krischer K 2015 Chimeras in globally coupled oscillatory systems: From ensembles of oscillators to spatially continuous media. \textit{Chaos} {\bf 25} 064401
		
		\bibitem{lr_glob} Miritello G,  Pluchino A, and Rapisarda A 2009 Phase transitions and chaos in long-range models of coupled oscillators  \textit{Euro. Phys. Lett} {\bf 85} 10007
		
		\bibitem{lr_lec} Dauxois  T,  Ruffo  S,  Arimondo  E and Wilkens M 2002 Dynamics and thermodynamics of systems with long range interactions,  \textit{Lect. Notes Phys.  (Springer)},	{\bf 602}
		
		\bibitem{s.gupta1}  Gupta S, and Ruffo S 2017 The world of long-range interactions: A bird’s eye view,  \textit{Int. J. Modern phys. A}  {\bf 32}  1741018 
		
		\bibitem{s.gupta2}  Bouchet F,  Gupta S 2010  Mukamel D,  Thermodynamics and dynamics of systems with long-range interactions \textit{Physica A: Statistical Mechanics and its Applications} {\bf 389} 4389-4405
		
		\bibitem{s.gupta3}  Nardini C,  Gupta S,  Ruffo S,  Dauxois T,  Bouchet F 2012 Kinetic theory for non-equilibrium stationary states in long-range interacting systems,  \textit{Journal of Statistical Mechanics: Theory and Experiment} {2012} L01002
		
		\bibitem{lr1} Campa A,  Dauxois T, and   Ruffo S 2009 Statistical mechanics and dynamics of solvable models with long-range interactions \textit{Physics Reports} {\bf 480} 57-159
		
		\bibitem{astro1}   Chavanis P H 2006  Phase transitions in self-gravitating systems \textit{Int. J. of Modern Phys. B} {\bf20} 3113-3198
		
		\bibitem{plasma}  Elskens Y,  Escande D F 2002 \textit{Microscopic Dynamics of Plasmas and Chaos}, IOP Publishing, Bristol. % Taylor & Francis  publisher
		
		\bibitem{hyd}  Miller J 1990 Statistical mechanics of Euler equations in two dimensions \textit{Phys. Rev. Lett.} {\bf 65} 2137
		
		
		\bibitem{ata1}  Mar\'odi M,  d’Ovidio F,  and  Vicsek T 2002 Synchronization of oscillators with long range interaction: Phase transition and anomalous finite size effects   \textit{Phys. Rev. E} {\bf 66} 011109
		
		\bibitem{ata2} Viana R L, Batista A M,  Batista C A S, de  Pontes J C A,  Silva F A S and Lopes S R 2012 Bursting synchronization in networks with long-range coupling mediated by a diffusing chemical substance \textit{Commun. Nonlinear Sci. Numer. Simulat.,} {\bf{17}} 2924-2942
		
		\bibitem{sync} Pikovsky, A., Rosenblum, M., and Kurths, J. 2001  Synchronization: A Universal Concept in Nonlinear Sciences (The city: Cambridge University Presss)
		
		\bibitem{syn0} Hammond C,  Bergman H, and Brown P 2007  Pathological synchronization in Parkinson's disease: networks, models and treatments \textit{Trends Neurosci. } {\bf 30} 357-364
		
		\bibitem{pop_syn1}  Popovych O V and  Tass P A 2014  Control of Abnormal Synchronization in Neurological Disorders \textit{Front Neurol.} {\bf  5} 268.
		
		\bibitem{syn2} Zheng Y, Wang G,Li K,  Bao G and Wang J 2014   Epileptic seizure prediction using phase synchronization based on bivariate empirical mode decomposition \textit{Clinical Neurophys.}  {\bf 125}  1104-1111
		
		\bibitem{awa_pras} Sabesan S,  Narayanan K,  Prasad A,  Spanias A,  Sackellares J C and  Iasemidis L D 2003 \textit{ Biomedical sciences instrumentation} {\bf 39} 129-135
		
		\bibitem{syn_dana} Bhowmick S K,  Ghosh D and  Dana S K 2011 Synchronization in counter-rotating oscillators \textit{Chaos} {\bf 21} 033118
		
		\bibitem{syn_kapi}  Koluda P,  Perlikowski P,  Czolczynski K and   Kapitaniak T 2014 Synchronization of two self-excited double pendula  \textit{The European Physical Journal Special Topics} {\bf 223} 613-629
		
		\bibitem{daido_anal1}  Daido H and  Nakanishi K 2006 Diffusion induced inhomogeneity in globally coupled oscillators: Swing-by mechanism 	\textit{Phys. Rev. Lett.} {\bf 96}  054101
		
		\bibitem{daido_anal2} Daido H and  Nakanishi K 2007 	Aging and clustering in globally coupled oscillators textit{Phys. Rev. E} {\bf 75}  056206
		
		
		
		
		\bibitem{main_prema}  Premalatha K, Chandrasekar V K,  Senthilvelan M, and   Lakshmanan M  2015 Impact of symmetry breaking in networks of globally coupled oscillators  \textit{Phys. Rev. E} {\bf 91} 052915
		
		
		
		
		
		\bibitem{sathya1}  Sathiyadevi K,   Karthiga S,  Chandrasekar V K,   Senthilkumar D V,  Lakshmanan M  2017 Spontaneous symmetry breaking due to the trade-off between attractive and repulsive couplings \textit{Phys. Rev. E} {\bf 95} 042301
		
		\bibitem{sathya2}  Sathiyadevi K,   Chandrasekar V K,   Senthilkumar D V,  Lakshmanan M   2018 Distinct collective states due to trade-off between attractive and repulsive couplings \textit{Phys. Rev. E} {\bf 97} 032207
		
		\bibitem{skdana}  Mishra A,  Hens C R,   Bose M,  Roy P K, and  Dana S K 2015
		Chimeralike states in a network of oscillators under attractive and repulsive global coupling  \textit{Phys. Rev. E} {\bf  92} 062920 
		
		
		
		\bibitem{stuart1}  Thompson M C  and Gal P L 2004 The Stuart–Landau model applied to wake transition revisited \textit{European Journal of Mechanics B/Fluids } {\bf{23}} 219-228
		
		\bibitem{stuart2} Moon J-Y,  Lee U,  Blain-Moraes S and Mashour G A 2015 General relationship of global topology, local dynamics, and directionality in large-scale brain networks \textit{PLoS Comput. Biol.} {\bf11} e1004225
		
		\bibitem{stuart3}Frasca M, Bergner A, Kurths J, and Fortuna L 2012 Bifurcation in a star-like network of Stuart-Landau oscillators. \textit{Int. J. of Bifur. and Chaos} {\bf 22} 1250173
		
		\bibitem{stu_ap}  Punetha N,  Varshney V,  Sahoo S, Saxena G,  Prasad A,  Ramaswamy R 2018 Dynamical effects of breaking rotational symmetry in counter-rotating Stuart-Landau oscillators \textit{Phys. Rev. E} {\bf 98} 022212
		
		
		\bibitem{sbre1}  Zakharova A,  Schneider I,  Yuliya Kyrychko Y, and Sch\"oll E 2013 Time delay control of symmetry-breaking primary and secondary oscillation death
		\textit{Euro. Phys. Lett}  104(5) 50004
		
		\bibitem{sbre2}  Schneider I,  Kapeller M,  Loos S, Zakharova A,  Fiedler B   Sch\o"ll E 2015 Stable and transient multicluster oscillation death in nonlocally coupled networks \textit{Phys. Rev. E}  {\bf 92} 052915 
		
		\bibitem{lr_range1}Xiang Li 2006 Phase synchronization in complex networks with decayed long-range interactions \textit{Physica D: Nonlinear Phenomena} {\bf 223} 242-247
		
		\bibitem{lr_range2}Iubini S,  Cintio P D, Lepri S,  Livi R,  and  Casetti L 2017 Heat transport in oscillator chains with long-range interactions coupled to thermal reservoirs 2018 \textit{Phys. Rev. E} {\bf 97} 032102 
		
		\bibitem{lr_range3} D Bagchi 2017 Thermal transport in the Fermi-Pasta-Ulam model with long-range interactions \textit{Phys. Rev. Lett.} {\bf 95} 032102
		
		\bibitem{lr_range4}	 Christodoulidi H,  Tsallis C, and  Bountis T 2014   Fermi-Pasta-Ulam model with long-range interactions: Dynamics
		and thermostatistics \textit{Euro. Phys. Lett} {\bf 108}  40006
		
		\bibitem{ss1}  Maistrenko Y,  Penkovsky B and  Rosenblum M 2014 Solitary state at the edge of synchrony in ensembles with attractive and repulsive interactions  \textit{Phys. Rev. E} {\bf 69} 060901(R)
		
		
		\bibitem{kapi1}  Jaros P,  Brezetsky S,  Levchenko R,  Dudkowski D,  Kapitaniak T,  Maistrenko Y 2018 Solitary states for coupled oscillators with inertia \textit{Chaos} {\bf 28} 011103
		
		\bibitem{si1}  Majhi S,  Muruganandam P,  Ferreira F,  Ghosh D, and  Dana S K 2018 
		Asymmetry in initial cluster size favors symmetry in a network of oscillators
		\textit{Chaos} {\bf 28} 081101 
		\bibitem{si2}  Gopal R,  Chandrasekar V K, Venkatesan A, and  Lakshmanan M  2014 Observation and characterization of chimera states in coupled dynamical systems with nonlocal coupling  \textit{Phys. Rev. E} {\bf 89} 052914
		
		\bibitem{vkc_anal}   Sheeba J H,  Chandrasekar V K, and   Lakshmanan M 2011 General coupled-nonlinear-oscillator model for event-related (de) synchronization \textit{Phys. Rev. E} {\bf 84} 036210
		
		\bibitem{gopal_anal}  Chandrasekar V K, Gopal R,  Venkatesan A,  and Lakshmanan M  2014 Mechanism for intensity-induced chimera states in globally coupled oscillators \textit{Phys. Rev. E} {\bf 90}  062913	
		
		
		
		\bibitem{msf1}  Pecora L M and   Carroll T L  1998 Master stability functions for synchronized coupled systems \textit{Phys. Rev. Lett.} {\bf 80}  2109
		
		
		\bibitem{msf2}  Sun J,  Bollt E M, and  Nishikawa T 2009 Master stability functions for coupled nearly identical dynamical systems \textit{ Euro. Phys. Lett.} {\bf 85}  60011
		
		\bibitem{msf3}  Acharyya S and  Amritkar R E 2015 Synchronization of nearly identical dynamical systems: Size instability  \textit{Phys. Rev. E} {\bf 92} 052902
		
		\bibitem{tbaner1}  Banerjee T,  Dutta P S,  Zakharova A and  Sch\"oll E  2016 Chimera patterns induced by distance-dependent power-law coupling in ecological networks	\textit{Phys. Rev. E} {\bf 94} 032206
		
		\bibitem{tbaner2}  Gupta A, Banerjee T and  Dutta P S 2017 Increased persistence via asynchrony in oscillating ecological populations with long-range interaction
		\textit{Phys. Rev. E} {\bf 96} 042202 	
		
	\end{thebibliography}
\end{document}